# A high efficiency architecture for cascaded Raman fiber lasers

V. R. Supradeepa,[1,*] Jeffrey W. Nichsolson,[1] Clifford E. Headley,[1] Man F. Yan,[1] Bera Palsdottir[2], and Dan Jakobsen[2]

[1]*OFS Laboratories, 19 Schoolhouse Road, Somerset, New Jersey 08873, USA*
[2] *OFS Fitel Denmark, Priorparken 680, 2605 Broendby, Denmark*
[*]*supradeepa@ofsoptics.com*

**Abstract:** We demonstrate a new high efficiency architecture for cascaded Raman fiber lasers based on a single pass cascaded amplifier configuration. Conversion is seeded at all intermediate Stokes wavelengths using a multi-wavelength seed source. A lower power Raman laser based on the conventional cascaded Raman resonator architecture provides a convenient seed source providing all the necessary wavelengths simultaneously. In this work we demonstrate a 1480nm laser pumped by an 1117nm Yb-doped fiber laser with maximum output power of 204W and conversion efficiency of 65% (quantum-limited efficiency is ~75%). We believe both the output power and conversion efficiency (relative to quantum-limited efficiency) are the highest reported for cascaded Raman fiber lasers.

**References and links**

1. S. G. Grubb, T. Erdogan, V. Mizrahi, T. Strasser, W. Y. Cheung, W. A. Reed, P. J. Lemaire, A. E. Miller, S. G. Kosinski, G. Nykolak, and P. C. Becker, "High power, 1.48 µm cascaded Raman laser in germanosilicate fibers," OSA Topic. Meeting, Optic. Amp. and Their Applications (1994).
2. S. K. Sim, H. C. Lim, L. W. Lee, L. C. Chia, R. F. Wu, I. Cristiani, M. Rini, and V. Degiorgio, "High-power cascaded Raman fibre laser using phosphosilicate fiber," Electron. Lett. **40**, 738-739 (2004).
3. Z. Xiong, N. Moore, Z. G. Li, and G. C. Lim, "10-W Raman fiber lasers at 1248 nm Using phosphosilicate fibers," J. Lightwave Technol. **21**, 2377-2381 (2003).
4. Y. Feng, L.R. Taylor, and D.B. Calia, "150 W highly-efficient Raman fiber laser," Opt. Express **17**, 23678-23683 (2009).
5. R. Vallee, E. Belanger, B. Dery, M. Bernier, and D. Faucher, "Highly efficient and High-power Raman fiber laser based on broadband chirped fiber Bragg gratings," J. Lightwave Technol. **24**, 5039-5043 (2006).
6. C. Headley and G. P. Agrawal, *"Raman Amplification in Fiber Optical Communication Systems"* (Elsevier, 2005).
7. D. Georgiev, V. P. Gapontsev, A. G. Dronov, M. Y. Vyatkin, A. B. Rulkov, S. V. Popov, and J. R. Taylor, "Watts-level frequency doubling of a narrow line linearly polarized Raman fiber laser to 589nm," Opt. Express **13**, 6772-6776 (2005).
8. J. C. Jasapara, M. J. Andrejco, A. D. Yablon, J. W. Nicholson, C. E. Headley, and D. J. DiGiovanni, "Picosecond pulse amplification in a core-pumped large-mode-area erbium fiber," Opt. Lett. **32**, 2429-2431 (2007).
9. J. W. Nicholson, J. M. Fini, A. M. DeSantolo, X. Liu, K. Feder, P. S. Westbrook, V. R. Supradeepa, E. Monberg, F. DiMarcello, R. Ortiz, C. Headley, and D. J. DiGiovanni, "Scaling the effective area of higher-order-mode erbium-doped fiber amplifiers," Opt. Express **20**, 24575-24584 (2012).
10. V. R. Supradeepa, J. W. Nicholson, and K. Feder, "Continuous wave Erbium-doped fiber laser with output power of >100 W at 1550 nm in-band core-pumped by a 1480nm Raman fiber laser," in CLEO: Science and Innovations, OSA Technical Digest (online) (Optical Society of America, 2012), paper CM2N.8.
11. Y. Emori, K. Tanaka, C. Headley, and A. Fujisaki, "High-power cascaded Raman fiber laser with 41-W output power at 1480-nm band," in Conference on Lasers and Electro-Optics/Quantum Electronics and Laser Science Conference and Photonic Applications Systems Technologies, OSA Technical Digest Series (CD) (OSA, 2007), paper CFI2.
12. J. W. Nicholson, M. F. Yan, P. Wisk, J. Fleming, F. DiMarcello, E. Monberg, T. Taunay, C. Headley, and D. J. DiGiovanni, "Raman fiber laser with 81 W output power at 1480 nm," Opt. Lett. **35**, 3069-3071 (2010).

13. M. A. Arbore, Y. Zhou, G. Keaton, and T. Kane, "36dB gain in S-band EDFA with distributed ASE suppression," in *Optical Amplifiers and Their Applications*, J. Nagel, S. Namiki, and L. Spiekman, eds., Vol. 77 of OSA Trends in Optics and Photonics Series (Optical Society of America, 2002), paper PD4.
14. P. D. Dragic, "Suppression of first order stimulated Raman scattering in erbium-doped fiber laser based LIDAR transmitters through induced bending loss," Opt. Commun. **250**, 403-410 (2005).
15. J. Kim, P. Dupriez, C. Codemard, J. Nilsson, and J. K. Sahu, "Suppression of stimulated Raman scattering in a high power Yb-doped fiber amplifier using a W-type core with fundamental mode cut-off," Opt. Express **14**, 5103-5113 (2006).
16. V. R. Supradeepa, J. W. Nicholson, C. E. Headley, Y. Lee, B. Palsdottir, and D. Jakobsen, "Cascaded Raman fiber faser at 1480nm with output power of 104W," no. 8237-48, SPIE photonics west 2012.
17. Y. Jeong, S. Yoo, C. A. Codemard, J. Nilsson, J. K. Sahu, D. N. Payne, R. Horley, P. W. Turner, L. M. B. Hickey, A. Harker, M. Lovelady, and A. Piper, "Erbium:ytterbium codoped large-core fiber laser with 297 W continuous-wave output power," IEEE J. Sel. Top. Quantum Electron. **13**, 573–579 (2007).
18. V. Kuhn, D. Kracht, J. Neumann, and P. Weßels, "Er-doped photonic crystal fiber amplifier with 70 W of output power," Opt. Lett. **36**, 3030-3032 (2011).
19. W. A. Reed, A. J. Stentz, and T. A. Strasser. "Article comprising a cascaded raman fiber laser," U.S. Patent 5,815,518 (1998).
20. C. Headley and G. Agrawal, "*Raman amplification in fiber optical communication systems,*" (Academic Press, 2005).
21. M. Rini, I. Cristiani, and V. Degiorgio, "Numerical modeling and optimization of cascaded CW Raman fiber lasers," IEEE J. Quantum Electron. **36**, 1117–1122 (2000).
22. S. D. Jackson and P. H. Muir, "Theory and numerical simulation of nth-order cascaded Raman fiber lasers," J. Opt. Soc. Am. B **18**, 1297-1306 (2001).
23. S. B. Papernyi, V. I. Karpov, and W. R. L. Clements, "Third-order cascaded Raman amplification," in Optical Fiber Communication Conference (OFC) 2002, FB4-1.IEEE, (2002).
24. S. Papernyi, V. Karpov, and W. Clements, "Cascaded pumping system and method for producing distributed Raman amplification in optical fiber telecommunication systems," U. S. Patent 6,480,326 (2002).
25. http://ofscatalog.specialityphotonics.com/category/high-power-products-and-cladding-pumped-fibers.
26. Y. Jeong, J. Sahu, D. Payne, and J. Nilsson, "Ytterbium-doped large-core fiber laser with 1.36 kW continuous-wave output power," Opt. Express **12**, 6088-6092 (2004).

## 1. Introduction

Cascaded Raman fiber lasers provide a convenient way to obtain high powers at wavelengths which may not be accessible through rare-earth doped fiber lasers [1-7]. The principle is to wavelength convert the output of a rare-earth doped fiber laser to the required output wavelength using a series of Raman Stokes shifts. Conventionally, wavelength conversion over two or more Stokes shifts is performed through the use of a cascaded Raman resonator (as shown schematically in Fig. 1(a)). It is comprised of nested cavities at each of the intermediate wavelengths made with fiber Bragg gratings (referred to as the Raman input and output grating sets) and a low effective area (high nonlinearity) fiber (Raman fiber). Each intermediate wavelength in the resonator is chosen to be close to the peak of the Raman gain of the wavelength preceding it. A low reflectivity output coupler terminates the wavelength conversion. At the output most of the light is at the desired final wavelength with small fractions at the intermediate wavelengths.

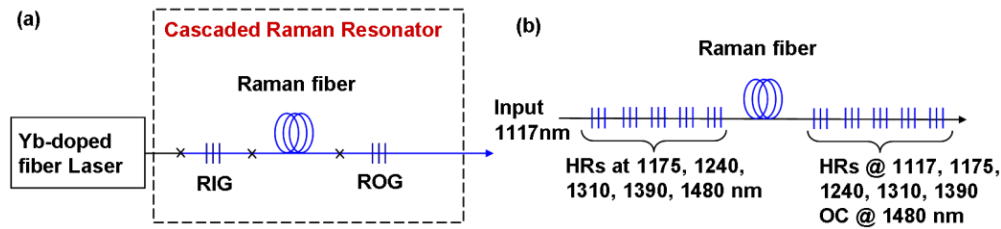

Fig. 1 (a) Schematic of a cascaded Raman laser, RIG – Raman input grating set, ROG – Raman output grating set, (b) Components of a cascaded Raman resonator converting 1117nm input to 1480nm output, HR – High reflectivity grating (> 99%), OC – output coupler, low reflectivity (< 10%) grating

Figure 1(b) shows an implementation of the cascaded Raman resonator performing five Stokes shifts from 1117nm to 1480nm. High power sources at 1.5 micron provide significantly higher eye safety than at the Yb wavelength region which is attractive for a variety of high power applications like material processing. Another interesting application for high power 1.5micron Raman lasers utilizes their ability to emit at the in-band absorption region of Erbium doped media.  They can provide convenient high brightness and low quantum defect pump sources for Erbium-doped fiber amplifiers resulting in high efficiency performance and low thermal load. Shorter amplifier lengths made possible with high brightness pumping results in reduced non-linearity in the amplifiers, compared to cladding pumping with multi-mode 9xx diode lasers. This is particularly attractive for pulsed or single frequency amplifiers. This method has been used to pump large-mode area (LMA) Er-doped fiber amplifiers [8], higher-order mode (HOM), Er-doped fiber amplifiers [9] and conventional Er fibers at high powers [10]. Due to high transparency of the earth's atmosphere at 1550nm, such sources are attractive for free space applications like LIDAR and directed energy.

Power scaling and enhancement of efficiency of Raman lasers has been a subject of current interest. The wavelength conversion efficiency in cascaded Raman lasers has been quite low compared to the quantum limited efficiency [1-7]. For example, in the case of 1117nm to 1480nm, the conversion efficiency has been < 48% while the quantum limited conversion efficiency is 75% [1]. This number dropped further to < 40% in early demonstrations for power scaling for a Raman laser with 40W at 1480nm [11], due to unwanted scattering of the 1480nm light to the next Stokes wavelength at 1590nm. This resulted in the use of short and suboptimal lengths of Raman fiber reducing the efficiency. This issue was overcome in [12] by the use of filter fibers [13-15] with enhanced losses at longer wavelengths which suppress further scattering of the output light. Fig. 2 shows the transmission function of such a filter fiber with a cut-off close to 1500nm for use in a 1480nm laser. This fiber was based on a W-shaped index profile which can create enhanced tunneling loss out of the fiber core for wavelengths longer than the cut-off wavelength [12]. More than 80W in [12] and 100W in [16] was demonstrated incorporating the filter fiber with conversion efficiency from 1117 nm to 1480 nm of ~48%.

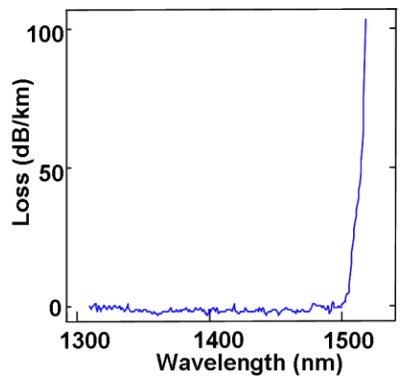

Fig. 2 Transmission function of the filter fiber used in the 1480 nm laser.

Here we propose a new architecture which achieves a significant enhancement in conversion efficiency compared to the previous architecture. For the case of 1117nm to 1480nm, we achieve a 65% conversion efficiency (for a quantum limited value of 75%) which significantly improves upon the 48% previously achieved for the same wavelengths. We demonstrate this new architecture through a 1480nm Raman laser which has a maximum output power of 204W. We believe both the output power and conversion efficiency (relative to quantum-limited efficiency) are the highest reported for cascaded Raman fiber lasers. The net optical to optical efficiency from 975nm pump diodes to 1480nm taking into account the efficiency of the Yb-doped fiber laser is ~43%.

The primary competing technologies for high power 1.5 micron fiber lasers pumped at 975nm (where mature, high power diode technology is available) are cladding pumped ErYb codoped fibers [17] and cladding pumped Er fibers [18]. Apart from parasitic lasing issues and beam quality problems, the efficiencies achieved at comparable power levels were significantly smaller than demonstrated here. Furthermore, the output power in our system was limited by total pump power and significant power scaling (limited only by thermal considerations and laser damage threshold) is possible. From the results demonstrated here we believe this is a very efficient and scalable approach to high power fiber lasers in the 1.5micron wavelength region.

## 2. Architecture

The primary sources responsible for the reduced efficiency in conventional cascaded Raman lasers can be identified as-

1. Transmission loss in the Raman input grating set and output grating set [19]

2. Two intra-cavity splices between low effective area (possibly dissimilar) fibers constituting the grating sets and the Raman gain fiber.

3. Linear loss in the Raman fiber.

4. Enhanced backward and forward light at the intermediate Stokes wavelengths due to their bandwidth being higher than the grating bandwidths [20-22].

5. Splice loss between the Yb-doped fiber laser output and the low effective area Raman fiber.

A number of loss components are associated with the cascaded Raman resonator assembly. Here we intend to eliminate the cascaded Raman resonator and use a single pass cascaded amplifier scheme. At higher powers this is expected to work very well as long as it is seeded at all the intermediate Stokes wavelengths with sufficient power. Physically, the seed powers at all the intermediate wavelengths are essential since they reduce the gain requirement, provide wavelength selectivity and preferential forward Raman scattering. The idea of using a pump separated by more than one Stokes shift from the signal with the wavelength conversion mediated by intermediate wavelengths has been used previously used in optical communications for distributed Raman amplifiers [23-24].

Two key ingredients are necessary to make the single pass cascaded amplifier feasible. Firstly, a simple multi-wavelength source which can simultaneously provide sufficient powers at all the intermediate wavelengths. A lower power conventional cascaded Raman laser lends itself ideally for this purpose. Light present at the output at all the intermediate Stokes wavelengths provides sufficient seed power at the exact required wavelengths. Secondly, scattering of the output wavelength to the next Stokes order can be further enhanced in a single pass configuration. The use of Raman filter fiber eliminates this problem and provides an ideal technique to terminate the cascade of wavelength conversion. Fig. 3(a) shows the experimental setup based on the new architecture. A high power Yb-doped fiber laser is combined with a lower power Raman seed laser. In our system, the Yb-doped fiber laser is emitting at 1117nm and the power is combined using an 1117/1480nm fused fiber WDM. This is then sent through ~50m of Raman filter fiber with an effective area of ~15 $\mu m^2$ and cut-off at 1500nm as shown in fig 2. The length of the Raman filter fiber was optimized experimentally and the obtained value was verified through numerical simulations. The high power Yb-doped fiber laser is based on a MOPA architecture with standard Yb125 fiber from OFS [25] and pumped with 18, nominally 25W, 975nm diodes. The laser can provide a maximum of 295W @ 1117nm limited by pump power. The conversion efficiency from 975nm to 1117nm is ~66%. Considering that slope efficiencies of 80% [26] have been reported, there is significant room for improvement in our overall system efficiency. The low power cascaded Raman laser has a nominal output power of 10W @ 1480nm and a total output power (including power at all the intermediate Stokes orders) of 12W. Fig. 3(b) shows the spectrum measured at the output with only the seed laser turned on. All the intermediate Stokes wavelengths necessary to mediate the cascaded conversion are seen.

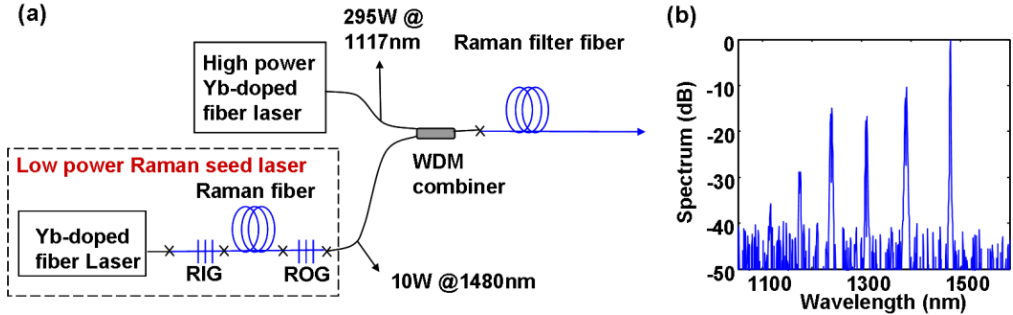

Fig. 3 (a) Schematic of the new cascaded Raman laser, RIG, ROG – Raman input and output grating sets (b) Spectrum measured at the output with only the seed source turned on

## 3. Experimental results

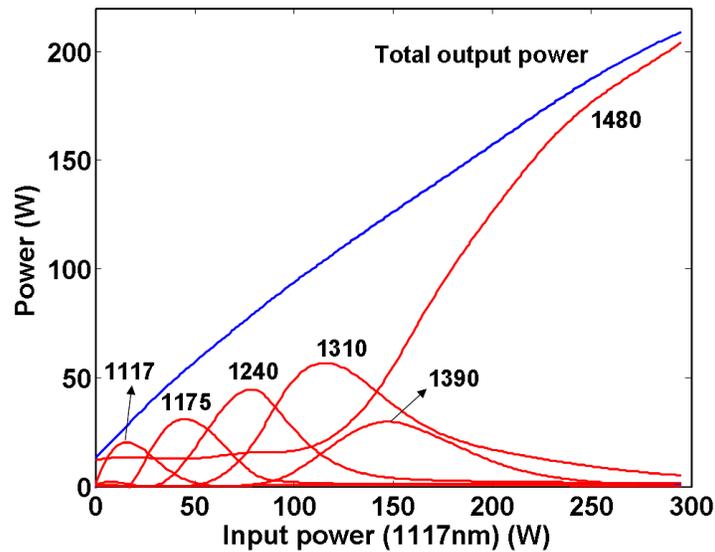

Fig. 4 Plot of total output power and components at each Stokes wavelength as a function of input power at 1117nm

Figure 4 shows the total output power and components at each Stokes wavelength measured as a function of input power at 1117nm to the cascaded amplifier. Interpolation has been used between the measured data points to better represent the evolution of powers. We clearly observe a progressive growth and decay of all the intermediate Stokes components with increasing power. A rapid growth of the final output wavelength is seen beyond a power threshold. The filter fiber ensures there is no further conversion of the output wavelength. At maximum power, we see a high degree of wavelength conversion with most of the output power being in the final, 1480nm component. The initial offset of the 1480nm component (and the total power) is due to the seed source. We also observe interesting behavior with the penultimate Stokes component (1390nm) which is unlike the previous ones. The presence of significant power already at the output wavelength (1480nm) manifests as additional loss through Stimulated Raman scattering for the 1390nm component. This creates a more complicated growth and decay condition. This behavior can become the limiting factor on power of the seed source. If the 1480nm component is too powerful, it can prematurely deplete the 1390nm component suppressing the power transfer from 1310nm. This can result in incomplete wavelength conversion and reduced efficiency. These aspects will be explored in more detail in future work.

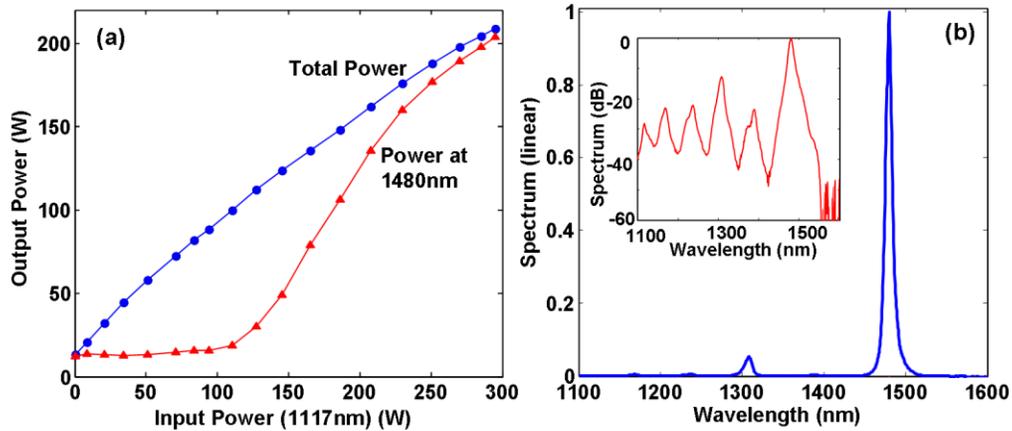

Fig. 5 (a) Plot of total output power and output power at 1480 nm as a function of input power at 1117 nm, (b) Spectrum of the output in linear and log scale (inset) at maximum power

Figure 5(a) shows the total output power and the 1480 nm component as a function of input power at 1117 nm (data points represented by circles and triangles are measurements). We have a purely single-moded output power of ~204 W at 1480 nm for a total input power of ~315 W (including the Yb fiber laser used in the seed source) with a total conversion efficiency of ~65% (1117nm to 1480nm). This is a significant enhancement in efficiency compared to previous work of ~48% [12, 16] being much closer to the quantum limited conversion efficiency of 75%. Taking into account the conversion efficiency of the Yb-doped fiber laser, we have an optical to optical conversion efficiency from 975 nm pump to 1480 nm signal of ~43%. This can be further increased by improving the Yb laser efficiency. Fig. 5(b) shows the measured output spectrum at full power (log scale in the inset). More than 95% of the power is in the 1480 nm band indicating the high level of wavelength conversion. No light is seen at the next Stokes order of 1590nm indicating the excellent suppression achieved by the Raman filter fiber.

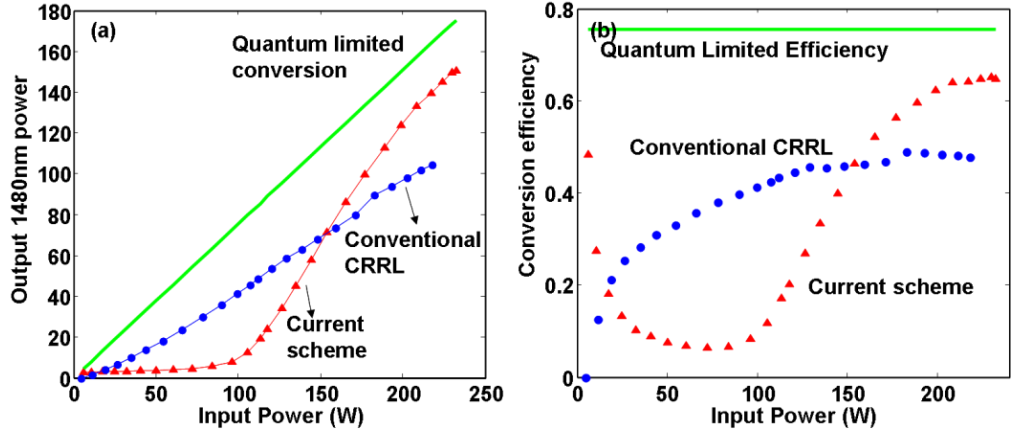

Fig. 6 (a) Comparison of output power at 1480nm as a function of input power at 1117nm between the new and conventional architectures with the quantum limited conversion. (b) Comparison of conversion efficiencies

To illustrate the enhancement in efficiency compared to the conventional architecture, we compare the results obtained with the current scheme to our previous results in [16] where 104W at 1480nm was obtained. To obtain optimal efficiencies at lower maximum input powers, the length was the Raman filter fiber was adjusted to ~65m. The power from the seed source in this experiment is ~3W at 1480nm and 4W in total. Fig 6(a) compares the output power at 1480nm for the two architectures together with the quantum-limited efficiency. At lower powers, power from the new scheme is low since the cascade is not driven all the way to 1480nm. However, once the threshold is reached, the power rapidly grows and at full power is significantly higher than that from the conventional cascaded Raman laser. At the maximum power point for the conventional Raman laser, the new scheme provides 40% more power. Fig. 6(b) shows the corresponding conversion efficiencies. At low power, the current scheme's output is dominated by the seed source and has the corresponding ~48% conversion efficiency. With increasing powers, but prior to threshold, this efficiency drops, but quickly recovers beyond that and achieves a maximum of ~65%. It is necessary to point out here that the efficiency of conventional cascaded Raman laser is power dependent as well (the optimal length of Raman fiber in the resonator needs to be modified) and the results plotted above from [16] achieve close to maximum efficiency at the highest power levels.

     Simulation studies to understand the optimal operating conditions for the new architecture as a function of input power, seed power, amplifier length etc are in progress. It is interesting however to point out that the difference in efficiency in the current scheme from the quantum limited efficiency can be mostly accounted for. The primary sources are splice loss between the output of the Yb-fiber to the much smaller Raman filter fiber (~5%) and residual power at all the intermediate Stokes wavelengths (~5%). This indicates that by reducing the splice loss through further optimization or novel splicing methods and better wavelength conversion through engineering the power and spectrum of the seed source, we should be able to further enhance the conversion efficiency.

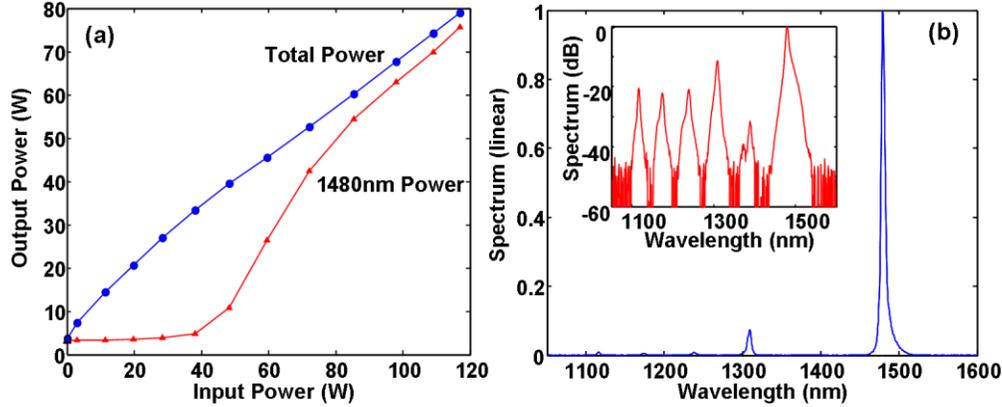

Fig. 7 A lower power demonstration – (a) Output power and 1480 nm component for a 75W Raman laser as a function of input power at 1117nm (b) Spectrum obtained at full power in linear and log scale.

To operate the new architecture at lower powers, a longer length of Raman filter fiber is necessary. However, the losses in these fibers are low (< 1dB/km) and the efficiency reduction by going to longer fibers is quite small. To demonstrate this, Fig. 7(a) shows the total output power and the 1480nm component of a system which provides ~75W at 1480nm for an input signal of 116W. In this experiment, the input source is a single stage Yb-doped fiber laser with a conversion efficiency of ~72%. The seed source used has an output power of 3W at 1480nm and 4W in total. The Raman filter fiber length has been increased to 150m (3 times that of the 200W result). Conversion efficiency from 1117nm to 1480nm in this case is slightly lower at 62%, due to increased linear loss in the Raman fiber, but still significantly higher than that of the conventional architecture. Fig. 7(b) shows the spectrum at full power in linear and log scale indicating a high level of wavelength conversion. The optical to optical efficiency from 975nm to 1480nm in this experiment is higher at ~45%. The comparative increase in total efficiency from the previous results in this paper is due to enhanced efficiency of the Yb oscillator. Comparing this work to the result in [12] where 81W at 1480nm was obtained, the net conversion efficiency has been enhanced from 32% to 45% at similar power levels.

## 4. Summary

In summary, we have demonstrated a new architecture for cascaded Raman lasers which achieves a significant enhancement in conversion efficiency compared to previous schemes. The architecture is based on a single pass cascaded Raman amplifier approach mediated at all intermediate Stokes wavelengths with a multi-wavelength seed source. A low power conventional cascaded Raman laser provides us with a convenient seed. In this work we demonstrated a 1480nm Raman laser pumped by an 1117nm Yb-doped fiber laser with output power of 204W. The output power achieved was only limited by available input power and the architecture has significant scaling potential. The conversion efficiencies were 65% from 1117 to 1480nm (for a quantum limited value of 75%) and 43% optical to optical from 975nm pump to 1480nm. We believe both the output power and conversion efficiency (relative to quantum limited efficiency) are the highest reported for cascaded Raman lasers. To demonstrate enhancement in efficiency at lower operating powers, we also demonstrated a 75W, 1480nm Raman laser. Future work will involve further power scaling efforts and simulation studies to better understand the relations between various system parameters and to obtain the optimal operating conditions. Obtaining pulsed Raman lasers through the use of pulsed input fiber lasers together with CW or pulsed seed lasers is also an attractive approach to follow.

Specifically for high power 1.5micron fiber lasers, the optical to optical efficiency from multi-mode 975nm pumps to 1480 nm laser demonstrated here (43%) is significantly higher than competing technologies at similar power levels based on cladding pumped Er and ErYb co-doped fibers. In this comparison, we do have to account for the difference in quantum defect between 975nm to 1480nm (this work) or to 1550nm (Er, ErYb). However this difference is quite small (~3%) and not significant compared to efficiency enhancements demonstrated here. We believe this is the most efficient and scalable approach to high power fiber lasers in the 1.5micron wavelength region. Another advantage specific to Raman lasers is their ability to act as high power, high brightness pump sources for pumping Er-doped media (like core-pumping large mode area Er-doped fiber amplifiers). This is an attractive option to reduce nonlinearity in high peak power pulsed amplifiers and single frequency amplifiers.